\documentclass{article}
\usepackage{amssymb}
\usepackage{amsmath}
\usepackage{amsfonts}
\usepackage{sw20ssur}
\usepackage{cite}
\usepackage{afterpage}

\input{tcilatex}
\begin{document}

\title{Constructing quantum games from a system of Bell's inequalities}
\author{Azhar Iqbal$^{\text{a,b}}$ and Derek Abbott$^{\text{a}}$ \\
%EndAName
$^{\text{a}}${\small School of Electrical \& Electronic Engineering, The
University of Adelaide, SA 5005, Australia.}\\
$^{\text{b}}${\small Centre for Advanced Mathematics and Physics,} {\small %
National University of Sciences \& Technology,}\\
{\small Peshawar Road, Rawalpindi, Pakistan.}}
\maketitle

\begin{abstract}
We report constructing quantum games directly from a system of Bell's
inequalities using Arthur Fine's analysis published in early 1980s. This
analysis showed that such a system of inequalities forms a set of both
necessary and sufficient conditions required to find a joint distribution
function compatible with a given set of joint probabilities, in terms of
which the system of Bell's inequalities is usually expressed. Using the
setting of a quantum correlation experiment for playing a quantum game, and
considering the examples of Prisoners' Dilemma and Matching Pennies, we
argue that this approach towards constructing quantum games addresses some
of their well known criticisms.
\end{abstract}


\section{Introduction}

Recent years have witnessed a growing interest in the area of quantum games~%
\cite%
{Mermin,Mermin1,MeyerDavid,MeyerDavid1,EWL,Vaidman,Brandt,BenjaminHayden,BenjaminHaydenC,EnkPike,Johnson,Johnson1,MarinattoWeber,IqbalToor1,Du,DuLi,IqbalToor2,Piotrowski, IqbalToor3,Piotrowski1,FlitneyAbbottA,IqbalToor4,Shimamura1,Shimamura2,FlitneyAbbottC,Mendes,CheonTsutsui,NawazToor, Shimamura,IqbalToor5,IchikawaTsutsui,CheonIqbal,IqbalWeigert,Ozdemir,IqbalCheon,Ramzan, FlitneyHollenberg1,Shimamura3,IqbalCheonM,Bleiler1,IqbalCheonChapter,Ichikawa,IqbalCheonAbbott,IqbalAbbott,FlitneyRecent,Faisal,Bleiler2}%
. It seems that there is now an agreement that construction of a quantum
game can be achieved along several possible routes \cite%
{MeyerDavid,EWL,MarinattoWeber,Piotrowski,CheonTsutsui,IqbalCheon,Bleiler1,Bleiler2}%
. The unifying idea underpinning these routes appears to consist of
establishing a link between the classical feature of a physical system,
shared by participating players and that facilitates the physical
implementation of the game, and the classical game, so that a classical game
results because of those features. It turns out that this link takes the
form of constraints placed on the (statistical) properties of the shared
physical system. A quantum game is then obtained by replacing the classical
features of the shared physical system with the quantum ones, while
retaining the mentioned link. Subsequently, one investigates the impact the
quantum feature(s) may have on the solution/outcome of the game.

As the mentioned link between the classical features of shared physical
system and the resulting classical game can be established in several
possible ways, there can exist many routes in obtaining a quantum game. In
the context of non-cooperative games this results in an interesting
situation that one route to quantize a game can provide a new quantum
mechanical Nash equilibrium (NE\footnote{%
In the rest of this paper we use NE to mean Nash Equilibrium or Nash
Equilibria. The correct meaning is to be judged from the context.}) that is
different from the one which another quantization route provides. In
Bleiler's words \cite{Bleiler1} comparing one route to another is similar to
comparing apples and oranges.

For instance, in Eisert et al.'s scheme \cite{EWL} for quantizing a game,
the classical game corresponds to an initial product quantum state, which
the players share for its local unitary manipulation and its later
measurement. In Marinatto and Weber's (MW) quantization scheme \cite%
{MarinattoWeber} the classical game corresponds to the product initial state 
$\left\vert 00\right\rangle $, as a pure state is forwarded to player for
its local manipulation by the identity $\hat{I}$ and Pauli $\hat{\sigma}_{x}$
operators before the final state is measured.

The link that Eisert et al.'s quantization scheme establishes between the
classical game and a product initial state (representing the `classicality'
of shared physical system) does not seem entirely convincing. It is because
a product initial state leads to a classical game but the classical game can
also result when players locally maneuver a maximally entangled state with
special unitary actions. Likewise, the relationship which the MW's scheme
establishes between the classical game and the product initial state $%
\left\vert 00\right\rangle $ does not appear convincing as a quantum game
can correspond to a pure initial state that does not violate Bell's
inequality.

It is true that in these quantization schemes a classical game is embedded
in the quantum game, but the respective embeddings do not ensure that a
quantum game results only when relevant Bell's inequalities are violated.
Along with these observations, Benjamin and Hayden \cite{BenjaminHaydenC}
pointed out that in the Eisert et al.'s quantization scheme players' unitary
actions are arbitrarily restricted and are not even closed under
composition. Also, referring to the same quantization formalism, Flitney and
Hollenberg \cite{FlitneyHollenberg1} pointed out that the new NE and the
classical-quantum transitions that occur are simply an artifact of the
particular strategy space chosen. Also, Enk and Pike's \cite{EnkPike}
remarked that as the scheme involves players having access to strategy sets
that are not available to them in the classical game, it makes sense to
equate the quantum game to an extended classical game constructed by adding
extra pure strategies in the game matrix.

In an effort to reply to these observations we proposed \cite%
{IqbalWeigert,IqbalCheon,IqbalCheonAbbott,IqbalAbbott} the setting of a
quantum correlation experiment \cite{EPR,Bohm,Bell,Aspect,Peres,Cereceda}
for playing a quantum game. The quantum game is played between two remotely
located agents/observers Alice and Bob who can perform measurements on parts
of a particle that has disintegrated into two. Alice and Bob each are given
two directions in which measurements are performed. In a run, each agent
performs a measurement along one of the two directions, whose outcome is a
dichotomic variable. A players' strategy is the probability distribution of
choosing between the two available directions.

A quantum correlation experiment seems to provide a natural setting for a
two-player two-strategy ($2\times 2$) quantum game. It is because a player
has to decide a pure strategy in each run and the probability distribution
over pure strategies defines a player's (mixed) strategy that is definable
over many runs. Referring to the quantum correlation experiment, we can
identify the agents Alice and Bob as players and assign the two available
directions to correspond to a player's pure strategies. Players' payoffs are
then expressed in terms of the joint probabilities relevant to the shared
physical systems, their moves or strategies, and the entries in the matrix
that defines the game.

In the present paper we present a new approach for construction of quantum
games that, once again, uses the setting of quantum correlation experiments.
We propose that a quantum game corresponding to a classical game should
satisfy the following two requirements: a) so as to avoid Enk and Pike type
argumentation \cite{EnkPike} the strategy sets available to the players are
identical in both the quantum and the corresponding classical game, b) the
quantization procedure should establish a convincing relation between the
classicality of the shared physical system, as it is expressed by a relevant
system of Bell's inequalities \cite{Busch}, and the classical game. We show
that when the classicality of a shared physical system is defined \cite%
{Busch} in terms of a system of Bell's inequalities it allows us to
establish such a relationship. We refer, in this connection, to the results
reported by Fine \cite{Fine} in early 1980s and build up our arguments on
them. These results are known to be significant with reference to joint
distributions, quantum correlations, and a system of Bell's inequalities,
all of which are relevant to the setting of quantum correlation experiments
that we use to play quantum games.

\section{Two-player two-strategy games}

Consider a two-player two-strategy game

\begin{equation}
\begin{array}{c}
\text{Alice}%
\end{array}%
\begin{array}{c}
S_{1} \\ 
S_{2}%
\end{array}%
\overset{\overset{%
\begin{array}{c}
\text{Bob}%
\end{array}%
}{%
\begin{array}{cccc}
S_{1}^{\prime } &  &  & S_{2}^{\prime }%
\end{array}%
}}{\left[ 
\begin{array}{cc}
(a_{1},b_{1}) & (a_{2},b_{2}) \\ 
(a_{3},b_{3}) & (a_{4},b_{4})%
\end{array}%
\right] },  \label{matrix}
\end{equation}%
in which $S_{1,2}$ and $S_{1,2}^{\prime }$ are Alice's and Bob's pure
strategies, respectively, whereas $a_{i}$ are Alice's and $b_{j}$ are Bob's
payoffs. For instance, we use $\Pi _{A}(S_{2},S_{1}^{\prime })$ to denote
Alice's payoff when she plays $S_{2}$ while Bob plays $S_{1}^{\prime }$,
which is $a_{3}$ from the Table (\ref{matrix}). In a mixed-strategy game, we
denote by $x$ the probability with which Alice chooses her pure strategy $%
S_{1}$. She then chooses $S_{2}$ with the probability $(1-x)$. Similarly, we
denote by $y$ the probability with which Bob chooses $S_{1}^{\prime }$. He
then chooses $S_{2}^{\prime }$ with the probability $(1-y)$. In this case we
write Alice's payoff by $\Pi _{A}(x,y)$ and Bob's payoff by $\Pi _{B}(x,y)$
i.e. the first entry in bracket is for Alice and the second for Bob. For a
symmetric game we have $a_{1}=b_{1},$ $a_{4}=b_{4},$ $a_{2}=b_{3},$ and $%
a_{3}=b_{2}$, for which one obtains $\Pi _{A}(x,y)=\Pi _{B}(y,x)$. The
inequalities

\begin{equation}
\Pi _{A}(x^{\star },y^{\star })-\Pi _{A}(x,y^{\star })\geqslant 0,\text{ }%
\Pi _{B}(x^{\star },y^{\star })-\Pi _{B}(x^{\star },y)\geqslant 0,
\label{NE}
\end{equation}%
describe that the strategy pair $(x^{\star },y^{\star })$ is a NE.

\section{Quantum games using correlation experiments}

In the setting of quantum correlation experiments~\cite{IqbalCheon}, which
we use to play a quantum version of the game (\ref{matrix}), players Alice
and Bob are located in space-time regions $R_{1}$ and $R_{2}$, respectively.
Ideally these regions are spacelike separated. Alice can perform
measurements on two bivalent observables (with values $\pm 1$) $A_{1}$ and $%
A_{2}$ in region $R_{1}$. Similarly, player Bob can perform measurements on
two bivalent observables (with values $\pm 1$) $B_{1}$ and $B_{2}$ in region 
$R_{2}$.

Referring to the matrix (\ref{matrix}) we make the associations $A_{1}\sim $ 
$S_{1},$ $A_{2}\sim $ $S_{2}$ and $B_{1}\sim $ $S_{1}^{\prime },$ $B_{2}\sim 
$ $S_{2}^{\prime }$ and take

\begin{equation}
\triangle _{1}=(a_{3}-a_{1}),\text{ }\triangle _{2}=(a_{4}-a_{2}),\text{ }%
\triangle _{3}=(\triangle _{2}-\triangle _{1}).  \label{DeltasDef}
\end{equation}%
We then consider the joint probabilities $P_{A_{i},B_{j}}$ (for $i=1,2$ and $%
j=1,2$) and denote $P(A_{1}B_{2})$, for example, the probability that both
the observable $A_{1}$ and $B_{2}$ take the value $+1$. Similarly, we denote 
$P(A_{1}\bar{B}_{2})$ for the joint probability when the observable $A_{1}$
takes the value $+1$ and the observable $B_{2}$ takes the value $-1$.

For the matrix game (\ref{matrix}) played in the setting of quantum
correlation experiments the payoff relations are expressed \cite{IqbalCheon}
as

\begin{equation}
\Pi _{A,B}(x,y)=\left[ 
\begin{array}{c}
x \\ 
1-x%
\end{array}%
\right] ^{T}\left[ 
\begin{array}{cc}
\Pi _{A,B}(S_{1},S_{1}^{\prime }) & \Pi _{A,B}(S_{1},S_{2}^{\prime }) \\ 
\Pi _{A,B}(S_{2},S_{1}^{\prime }) & \Pi _{A,B}(S_{2},S_{2}^{\prime })%
\end{array}%
\right] \left[ 
\begin{array}{c}
y \\ 
1-y%
\end{array}%
\right] ,  \label{payoffs}
\end{equation}%
where $T$ is transpose and subscripts $A$ and $B$ refer to Alice and Bob,
respectively. In (\ref{payoffs}) we define

\begin{eqnarray}
\Pi _{A,B}(S_{1},S_{1}^{\prime }) &=&(a,b)_{1}P(A_{1}B_{1})+(a,b)_{2}P(A_{1}%
\bar{B}_{1})+(a,b)_{3}P(\bar{A}_{1}B_{1})+(a,b)_{4}P(\bar{A}_{1}\bar{B}_{1}),
\notag \\
\Pi _{A,B}(S_{1},S_{2}^{\prime }) &=&(a,b)_{1}P(A_{1}B_{2})+(a,b)_{2}P(A_{1}%
\bar{B}_{2})+(a,b)_{3}P(\bar{A}_{1}B_{2})+(a,b)_{4}P(\bar{A}_{1}\bar{B}_{2}),
\notag \\
\Pi _{A,B}(S_{2},S_{1}^{\prime }) &=&(a,b)_{1}P(A_{2}B_{1})+(a,b)_{2}P(A_{2}%
\bar{B}_{1})+(a,b)_{3}P(\bar{A}_{2}B_{1})+(a,b)_{4}P(\bar{A}_{2}\bar{B}_{1}),
\notag \\
\Pi _{A,B}(S_{2},S_{2}^{\prime }) &=&(a,b)_{1}P(A_{2}B_{2})+(a,b)_{2}P(A_{2}%
\bar{B}_{2})+(a,b)_{3}P(\bar{A}_{2}B_{2})+(a,b)_{4}P(\bar{A}_{2}\bar{B}_{2}).
\notag \\
&&  \label{payoffs-parts}
\end{eqnarray}%
where $(a,b)_{2}$, for example, is shortened notation for $a_{2},b_{2}$%
---the entries in the matrix (\ref{matrix}).

Although players' payoffs (\ref{payoffs}) depend on the joint probabilities
corresponding to the shared physical system, the players' moves, represented
by $x$ and $y$, are independent of them. Players' moves in the quantum game
are classical in being a linear combination (with real \& normalized
coefficients) of the two choices available to each player. However, in
contrast to the situation in a classical game (in which the chosen
strategies directly determine the payoff entries in the payoff matrix), our
setting demands that players' payoffs not only depend on their moves but
also that these depend on what kind of physical system players share in
order to play the game. We achieve this by making the payoff relations to
depend also on the joint probabilities relevant to the shared physical
system and then ask whether a joint probability distribution exists. As for
a quantum mechanical shared physical system the joint probabilities can go
beyond the constraints permitted to classical joint probabilities, allowing
us to obtain our quantum game.

As the joint probabilities are normalized we have

\begin{eqnarray}
P(A_{1}B_{1})+P(A_{1}\bar{B}_{1})+P(\bar{A}_{1}B_{1})+P(\bar{A}_{1}\bar{B}%
_{1})=1, &&  \notag \\
P(A_{1}B_{2})+P(A_{1}\bar{B}_{2})+P(\bar{A}_{1}B_{2})+P(\bar{A}_{1}\bar{B}%
_{2})=1, &&  \notag \\
P(A_{2}B_{1})+P(A_{2}\bar{B}_{1})+P(\bar{A}_{2}B_{1})+P(\bar{A}_{2}\bar{B}%
_{1})=1, &&  \notag \\
P(A_{2}B_{2})+P(A_{2}\bar{B}_{2})+P(\bar{A}_{2}B_{2})+P(\bar{A}_{2}\bar{B}%
_{2})=1, &&  \label{normalization}
\end{eqnarray}%
and thus each one of the relations (\ref{payoffs-parts}) represents a
classical mixed strategy payoff. The causal communication constraint \cite%
{Cereceda} for the joint probabilities $P_{A_{i},B_{j}}$ (for $i=1,2$ and $%
j=1,2$) now states that

\begin{eqnarray}
P(A_{1}B_{1})+P(A_{1}\bar{B}_{1}) &=&P(A_{1}B_{2})+P(A_{1}\bar{B}_{2}),\text{
}P(A_{1}B_{1})+P(\bar{A}_{1}B_{1})=P(A_{2}B_{1})+P(\bar{A}_{2}B_{1}),  \notag
\\
P(A_{2}B_{1})+P(A_{2}\bar{B}_{1}) &=&P(A_{2}B_{2})+P(A_{2}\bar{B}_{2}),\text{
}P(A_{1}B_{2})+P(\bar{A}_{1}B_{2})=P(A_{2}B_{2})+P(\bar{A}_{2}B_{2}),  \notag
\\
P(\bar{A}_{1}B_{1})+P(\bar{A}_{1}\bar{B}_{1}) &=&P(\bar{A}_{1}B_{2})+P(\bar{A%
}_{1}\bar{B}_{2}),\text{ }P(\bar{A}_{2}B_{1})+P(\bar{A}_{2}\bar{B}_{1})=P(%
\bar{A}_{2}B_{2})+P(\bar{A}_{2}\bar{B}_{2}),  \notag \\
P(A_{1}\bar{B}_{1})+P(\bar{A}_{1}\bar{B}_{1}) &=&P(A_{2}\bar{B}_{1})+P(\bar{A%
}_{2}\bar{B}_{1}),\text{ }P(A_{1}\bar{B}_{2})+P(\bar{A}_{1}\bar{B}%
_{2})=P(A_{2}\bar{B}_{2})+P(\bar{A}_{2}\bar{B}_{2}).  \notag \\
&&  \label{locality}
\end{eqnarray}%
Using Eqs.~(\ref{normalization},\ref{locality}) it can be shown \cite%
{Cereceda} that $8$ out of $16$ joint probabilities $P_{A_{i},B_{j}}$ (for $%
i=1,2$ and $j=1,2$) can be eliminated.

With the payoff relations (\ref{payoffs}) the Nash inequalities for an
arbitrary pair of strategies $(x^{\star },y^{\star })$ are written as

\begin{eqnarray}
\Pi _{A}(x^{\star },y^{\star })-\Pi _{A}(x,y^{\star }) &=&(x^{\star
}-x)\{y^{\star
}\{a_{1}[P(A_{1}B_{1})-P(A_{1}B_{2})-P(A_{2}B_{1})+P(A_{2}B_{2})]+  \notag \\
&&a_{2}[P(A_{1}\bar{B}_{1})-P(A_{1}\bar{B}_{2})-P(A_{2}\bar{B}_{1})+P(A_{2}%
\bar{B}_{2})]+  \notag \\
&&a_{3}[P(\bar{A}_{1}B_{1})-P(\bar{A}_{1}B_{2})-P(\bar{A}_{2}B_{1})+P(\bar{A}%
_{2}B_{2})]+  \notag \\
&&a_{4}[P(\bar{A}_{1}\bar{B}_{1})-P(\bar{A}_{1}\bar{B}_{2})-P(\bar{A}_{2}%
\bar{B}_{1})+P(\bar{A}_{2}\bar{B}_{2})]\}+  \notag \\
&&\{a_{1}[P(A_{1}B_{2})-P(A_{2}B_{2})]+a_{2}[P(A_{1}\bar{B}_{2})-P(A_{2}\bar{%
B}_{2})]+  \notag \\
&&a_{3}[P(\bar{A}_{1}B_{2})-P(\bar{A}_{2}B_{2})]+a_{4}[P(\bar{A}_{1}\bar{B}%
_{2})-P(\bar{A}_{2}\bar{B}_{2})]\}\}\geq 0,  \label{NEx}
\end{eqnarray}

\begin{eqnarray}
\Pi _{B}(x^{\star },y^{\star })-\Pi _{B}(x^{\star },y) &=&(y^{\star
}-y)\{x^{\star
}\{b_{1}[P(A_{1}B_{1})-P(A_{1}B_{2})-P(A_{2}B_{1})+P(A_{2}B_{2})]+  \notag \\
&&b_{2}[P(A_{1}\bar{B}_{1})-P(A_{1}\bar{B}_{2})-P(A_{2}\bar{B}_{1})+P(A_{2}%
\bar{B}_{2})]+  \notag \\
&&b_{3}[P(\bar{A}_{1}B_{1})-P(\bar{A}_{1}B_{2})-P(\bar{A}_{2}B_{1})+P(\bar{A}%
_{2}B_{2})]+  \notag \\
&&b_{4}[P(\bar{A}_{1}\bar{B}_{1})-P(\bar{A}_{1}\bar{B}_{2})-P(\bar{A}_{2}%
\bar{B}_{1})+P(\bar{A}_{2}\bar{B}_{2})]\}+  \notag \\
&&\{b_{1}[P(A_{2}B_{1})-P(A_{2}B_{2})]+b_{2}[P(A_{2}\bar{B}_{1})-P(A_{2}\bar{%
B}_{2})]+  \notag \\
&&b_{3}[P(\bar{A}_{2}B_{1})-P(\bar{A}_{2}B_{2})]+b_{4}[P(\bar{A}_{2}\bar{B}%
_{1})-P(\bar{A}_{2}\bar{B}_{2})]\}\}\geq 0.  \label{NEy}
\end{eqnarray}

Notice that with respect to the joint probability distribution $%
P_{A_{1},A_{2},B_{1},B_{2}}$, if it exists, the given joints $%
P_{A_{i},B_{j}} $ (for $i=1,2$ and $j=1,2$) can be expressed as their
marginals. For instance

\begin{equation}
P(A_{2}\bar{B}_{1})=P(A_{1}A_{2}\bar{B}_{1}B_{2})+P(A_{1}A_{2}\bar{B}_{1}%
\bar{B}_{2})+P(\bar{A}_{1}A_{2}\bar{B}_{1}B_{2})+P(\bar{A}_{1}A_{2}\bar{B}%
_{1}\bar{B}_{2}).  \label{Marginals}
\end{equation}%
Similar expressions can be written for $P(A_{i}B_{j}),$ $P(\bar{A}%
_{i}B_{j}), $ and $P(\bar{A}_{i}\bar{B}_{j})$ for $i=1,2$ and $j=1,2$. In
the rest of this paper we refer to $P_{A_{i},B_{j}}$ as joint probabilities
and to $P_{A_{1},A_{2},B_{1},B_{2}}$ as the joint probability distribution.

\section{Fine's analysis}

At this stage we refer to a result reported in early 1980s by Arthur Fine 
\cite{Fine} stating that Bell's inequalities form both necessary and
sufficient conditions in order to find a joint probability distribution $%
P_{A_{1},A_{2},B_{1},B_{2}}$ whose marginals are the joint probabilities $%
P_{A_{i},B_{j}}$ (for $i=1,2$ and $j=1,2$). For the case when Bell's
inequalities hold, Fine describes how to find the probability distribution $%
P_{A_{1},A_{2},B_{1},B_{2}}$ from the joints probabilities $P_{A_{i},B_{j}}$%
, in terms of which the inequalities are usually expressed.

Fine presents two theorems, the first of which states that if $A,$ $B,$ $%
B^{\prime }$ are bivalent observables, each mapping into $\left\{
+1,-1\right\} $ with given joint distributions $P_{A,B},$ $P_{A,B^{\prime }}$
and $P_{B,B^{\prime }},$ then the necessary and sufficient condition for the
existence of a joint distribution $P_{A,B,B^{\prime }}$, compatible with the
given joints for the pairs, is the satisfaction of following system of
inequalities:

\begin{eqnarray}
P(A)+P(B)+P(B^{\prime }) &\leq &1+P(AB)+P(AB^{\prime })+P(BB^{\prime }), 
\notag \\
P(AB)+P(AB^{\prime }) &\leq &P(A)+P(BB^{\prime }),  \notag \\
P(AB)+P(BB^{\prime }) &\leq &P(B)+P(AB^{\prime }),  \notag \\
P(AB^{\prime })+P(BB^{\prime }) &\leq &P(B^{\prime })+P(AB),
\end{eqnarray}%
where $P(\cdot)$ denotes the probability that each enclosed observable takes
the value $+1$.

Fine's second theorem \cite{Fine} states that if $A_{1},$ $A_{2},$ $B_{1},$ $%
B_{2}$ are bivalent observables with joint distributions $P_{A_{i}},_{B_{j}}$
(for $i=1,2$ and $j=1,2$), then the necessary and sufficient condition for
there to exist a joint distribution $P_{A_{1},A_{2},B_{1},B_{2}}$ compatible
with the given joints is that the following system of Bell's inequalities is
satisfied:

\begin{equation}
-1\leq P(A_{i}B_{j})+P(A_{i}B_{j^{\prime }})+P(A_{i^{\prime }}B_{j^{\prime
}})-P(A_{i^{\prime }}B_{j})-P(A_{i})-P(B_{j^{\prime }})\leq 0,  \label{BIs}
\end{equation}%
for $i\neq i^{\prime }=1,2$ and $j\neq j^{\prime }=1,2$. The second theorem
becomes particularly relevant as it relates to the setting of quantum
correlation experiments that we use in this paper to play a quantum game.

By these theorems Fine finds the joint probability distribution $%
P_{A_{1},A_{2},B_{1},B_{2}}$ by letting $n=1,2$ and $m\neq k=1,2$ and by
setting

\begin{equation}
\gamma =\min \left\{
P(A_{n}B_{m})+P(B_{k})-P(A_{n}B_{k}),P(B_{m}),P(B_{k})\right\} .
\label{Gamma}
\end{equation}%
Afterwards, by defining $P(B_{1}B_{2})=\gamma $ Fine fills the rest of the
distribution by letting

\begin{eqnarray}
P(\bar{B}_{1}B_{2}) &=&P(B_{1})-\gamma ,  \notag \\
P(B_{1}\bar{B}_{2}) &=&P(B_{2})-\gamma ,  \notag \\
P(\bar{B}_{1}\bar{B}_{2}) &=&1-P(B_{1})-P(B_{2})+\gamma .
\end{eqnarray}%
Fine then defines two quantities $\alpha $ and $\beta $ as

\begin{equation}
\alpha =P(A_{1}B_{1}B_{2})=\gamma P(A_{1}),\text{ \ \ }\beta
=P(A_{2}B_{1}B_{2})=\gamma P(A_{2})  \label{AlphaBeta}
\end{equation}%
to find the distributions $P_{A_{1},B_{1},B_{2}}$ and $P_{A_{2},B_{1},B_{2}}$
as given below.

\begin{eqnarray}
P(A_{1}B_{1}\bar{B}_{2}) &=&P(A_{1}B_{1})-\alpha ,  \notag \\
P(A_{1}\bar{B}_{1}B_{2}) &=&P(A_{1}B_{2})-\alpha ,  \notag \\
P(A_{1}\bar{B}_{1}\bar{B}_{2})
&=&P(A_{1})-P(A_{1}B_{1})-P(A_{1}B_{2})+\alpha ,  \notag \\
P(\bar{A}_{1}B_{1}B_{2}) &=&P(B_{1}B_{2})-\alpha ,  \notag \\
P(\bar{A}_{1}B_{1}\bar{B}_{2})
&=&P(B_{1})-P(A_{1}B_{1})-P(B_{1}B_{2})+\alpha ,  \notag \\
P(\bar{A}_{1}\bar{B}_{1}B_{2})
&=&P(B_{2})-P(A_{1}B_{2})-P(B_{1}B_{2})+\alpha ,  \notag \\
P(\bar{A}_{1}\bar{B}_{1}\bar{B}_{2})
&=&1-P(A_{1})-P(B_{1})-P(B_{2})+P(A_{1}B_{1})+P(A_{1}B_{2})+P(B_{1}B_{2})-%
\alpha ,  \notag \\
&&  \label{Fine1}
\end{eqnarray}%
and

\begin{eqnarray}
P(A_{2}B_{1}\bar{B}_{2}) &=&P(A_{2}B_{1})-\beta ,  \notag \\
P(A_{2}\bar{B}_{1}B_{2}) &=&P(A_{2}B_{2})-\beta ,  \notag \\
P(A_{2}\bar{B}_{1}\bar{B}_{2}) &=&P(A_{2})-P(A_{2}B_{1})-P(A_{2}B_{2})+\beta
,  \notag \\
P(\bar{A}_{2}B_{1}B_{2}) &=&P(B_{1}B_{2})-\beta ,  \notag \\
P(\bar{A}_{2}B_{1}\bar{B}_{2}) &=&P(B_{1})-P(A_{2}B_{1})-P(B_{1}B_{2})+\beta
,  \notag \\
P(\bar{A}_{2}\bar{B}_{1}B_{2}) &=&P(B_{2})-P(A_{2}B_{2})-P(B_{1}B_{2})+\beta
,  \notag \\
P(\bar{A}_{2}\bar{B}_{1}\bar{B}_{2})
&=&1-P(A_{2})-P(B_{1})-P(B_{2})+P(A_{2}B_{1})+P(A_{2}B_{2})+P(B_{1}B_{2})-%
\beta ,  \notag \\
&&  \label{Fine2}
\end{eqnarray}%
from which the distribution $P_{A_{1},A_{2},B_{1},B_{2}}$ can easily be
found.

In the following, while using Fine's analysis, we analyze the quantum games
of Prisoners's Dilemma (PD) and Matching Pennies (MP) played in the setting
of quantum correlation experiments. We have selected these games because
both games have been analyzed earlier in Refs.~\cite{IqbalCheon,IqbalAbbott}
using the concept of non-factorizable joint probabilities. This will provide
us an opportunity to find how the outcomes of these quantum games compare
when the games are studied using the present approach built up on Fine's
analysis and the approach building up on the joint probabilities becoming
non-factorizable. Using the concept of non-factorizable joint probabilities,
the quantum PD game produces no new outcome over the classical one. Using
the same procedure for quantum MP game, however, results in new
non-classical NE when the players share a entangled state that maximally
violates the Clauser-Holt-Shimony-Horne (CHSH) sum of correlations.

A second reason to chose these games is that classically each of these two
games have only one NE---a pure one for PD and mixed one for MP. As it is
the case with the approach using non-factorizable joint probabilities, the
present approach, based on Fine's analysis, also uses constraints on joint
probabilities that are associated with a particular NE. The situation of
having a unique classical NE presents a more easily tractable case when we
are considering constraints on quantum mechanical joint probabilities,
relative to the case when multiple NE exist for a game and we have to
separately consider constraints on joint probabilities for each of them.

\section{Analysis of the quantum Prisoners' Dilemma game}

We now refer to the matrix (\ref{matrix}) and consider the PD game for which
we have $a_{3}>a_{1}>a_{4}>a_{2}$ and as it is a symmetric game we have

\begin{equation}
\left( 
\begin{array}{cc}
a_{1} & a_{2} \\ 
a_{3} & a_{4}%
\end{array}%
\right) ^{T}=\left( 
\begin{array}{cc}
b_{1} & b_{2} \\ 
b_{3} & b_{4}%
\end{array}%
\right) ,
\end{equation}%
where $T$ is for transpose. Referring to (\ref{DeltasDef}) we then have $%
\triangle _{1},\triangle _{2}>0$ and $\Delta _{1}=b_{2}-b_{1}$ and $\Delta
_{2}=b_{4}-b_{3}$. The strategy pair $(x^{\star },y^{\star })=(0,0)$ can be
shown to emerge as a NE for this game at which we have $\Pi _{A}(0,0)=a_{4}$
and $\Pi _{B}(0,0)=b_{4}$.

To consider the quantum version of this game played using the setting of
quantum correlation experiments, we consider the strategy pair $(x^{\star
},y^{\star })=(0,0)$ for which we define

\begin{equation}
\triangle _{x^{\star }=0}=\Pi _{A}(0,0)-\Pi _{A}(x,0),\text{ \ \ }\triangle
_{y^{\star }=0}=\Pi _{B}(0,0)-\Pi _{B}(0,y).
\end{equation}%
For this strategy pair, we insert from Eqs.~(\ref{Marginals}) into the
inequalities (\ref{NEx},\ref{NEy}) to obtain

\begin{eqnarray}
\triangle _{x^{\star }=0} &{\small =}&{\small x[}\triangle _{1}\left\{
P(A_{1}\bar{A}_{2}B_{1}B_{2})-P(\bar{A}_{1}A_{2}B_{1}B_{2})-P(\bar{A}%
_{1}A_{2}\bar{B}_{1}B_{2})+P(A_{1}\bar{A}_{2}\bar{B}_{1}B_{2})\right\} + 
\notag \\
&&\triangle _{2}\left\{ P(A_{1}\bar{A}_{2}B_{1}\bar{B}_{2})-P(\bar{A}%
_{1}A_{2}B_{1}\bar{B}_{2})-P(\bar{A}_{1}A_{2}\bar{B}_{1}\bar{B}_{2})+P(A_{1}%
\bar{A}_{2}\bar{B}_{1}\bar{B}_{2})\right\} ],  \notag \\
\triangle _{y^{\star }=0} &{\small =}&{\small y[}\triangle _{1}\left\{
P(A_{1}A_{2}B_{1}\bar{B}_{2})-P(A_{1}A_{2}\bar{B}_{1}B_{2})-P(\bar{A}%
_{1}A_{2}\bar{B}_{1}B_{2})+P(\bar{A}_{1}A_{2}B_{1}\bar{B}_{2})\right\} + 
\notag \\
&&\triangle _{2}\left\{ P(A_{1}\bar{A}_{2}B_{1}\bar{B}_{2})-P(A_{1}\bar{A}%
_{2}\bar{B}_{1}B_{2})-P(\bar{A}_{1}\bar{A}_{2}\bar{B}_{1}B_{2})+P(\bar{A}_{1}%
\bar{A}_{2}B_{1}\bar{B}_{2})\right\} ],  \label{NEs}
\end{eqnarray}%
that express $\triangle _{x^{\star }=0}$ and $\triangle _{y^{\star }=0}$ in
terms of the joint probability distribution $P_{A_{1},A_{2},B_{1},B_{2}}$.

We now demand that the strategy pair $(0,0)$ emerges as a NE when players
share a classical physical system. For this we impose the following two
requirements: When the system of Bell's inequalities (\ref{BIs})\ holds we
have: a) Nash inequalities for the strategy pair $(0,0)$ hold i.e. $%
\triangle _{x^{\star }=0}\geq 0$, $\triangle _{y^{\star }=0}\geq 0$ and b)
the payoffs for players Alice and Bob at the strategy pair $(0,0)$ are $%
a_{4} $\ and $b_{4}$, respectively. These requirements ensure that a
faithful realization of the original game exists within the quantum game
constructed using the setting of quantum correlation experiments.

A keen reader may ask here why the converse situation is not adopted i.e. to
require that if $\triangle _{x^{\star }=0}\geq 0$ and $\triangle _{y^{\star
}=0}\geq 0$ then the system of Bell's inequalities holds. Unfortunately,
this is not a valid requirement as the violation of Bell's inequalities does
not necessarily lead to a new non-classical NE, and we well may have $%
\triangle _{x^{\star }=0}\geq 0$ and $\triangle _{y^{\star }=0}\geq 0$ even
when the system of Bell's inequalities is violated.

These requirements remind us of Fine's second theorem, while Eqs.~(\ref%
{Fine1},\ref{Fine2}) allow us to find the joint probability distribution $%
P_{A_{1},A_{2},B_{1},B_{2}}$. Using Eqs.~(\ref{Fine1},\ref{Fine2}) we,
therefore, re-express the quantities $\triangle _{x^{\star }=0}$ and $%
\triangle _{y^{\star }=0}$ in Eqs.~(\ref{NEs}) in terms of the joint
probabilities $P_{A_{i},B_{j}}$ (for $i=1,2$ and $j=1,2$) to obtain%
\begin{gather}
P(A_{1}\bar{A}_{2}B_{1}B_{2})+P(A_{1}\bar{A}_{2}\bar{B}_{1}B_{2})-P(\bar{A}%
_{1}A_{2}B_{1}B_{2})-  \notag \\
P(\bar{A}_{1}A_{2}\bar{B}_{1}B_{2})=[P(A_{1})-P(A_{2})]P(B_{2}),  \notag \\
P(A_{1}\bar{A}_{2}B_{1}\bar{B}_{2})+P(A_{1}\bar{A}_{2}\bar{B}_{1}\bar{B}%
_{2})-P(\bar{A}_{1}A_{2}B_{1}\bar{B}_{2})-  \notag \\
P(\bar{A}_{1}A_{2}\bar{B}_{1}\bar{B}_{2})=[P(A_{1})-P(A_{2})][1-P(B_{2})], 
\notag \\
P(A_{1}A_{2}B_{1}\bar{B}_{2})+P(\bar{A}_{1}A_{2}B_{1}\bar{B}%
_{2})-P(A_{1}A_{2}\bar{B}_{1}B_{2})-  \notag \\
P(\bar{A}_{1}A_{2}\bar{B}_{1}B_{2})=[P(B_{1})-P(B_{2})]P(A_{2}),  \notag \\
P(A_{1}\bar{A}_{2}B_{1}\bar{B}_{2})+P(\bar{A}_{1}\bar{A}_{2}B_{1}\bar{B}%
_{2})-P(\bar{A}_{1}\bar{A}_{2}\bar{B}_{1}B_{2})-  \notag \\
P(A_{1}\bar{A}_{2}\bar{B}_{1}B_{2})=[P(B_{1})-P(B_{2})][1-P(A_{2})],
\end{gather}%
and the Nash inequalities for the strategy pair $(0,0)$ then take a simpler
form

\begin{gather}
\triangle _{x^{\star }=0}=x[P(A_{1})-P(A_{2})][(\triangle _{1}/\triangle
_{2}-1)P(B_{2})+1]\triangle _{2}\geq 0,  \notag \\
\triangle _{y^{\star }=0}=y[P(B_{1})-P(B_{2})][(\triangle _{1}/\triangle
_{2}-1)P(A_{2})+1]\triangle _{2}\geq 0,  \label{Constraints}
\end{gather}%
giving the first set of constraints on joint probabilities as

\begin{equation}
P(A_{1})\geq P(A_{2}),\text{ \ \ }P(B_{1})\geq P(B_{2}).
\label{ConstraintsB}
\end{equation}

Similarly, below we translate the requirement b) into the second set of
constraints on joint probabilities. Consider the relations (\ref{payoffs})\
to find players' payoffs at the strategy pair $(0,0)$ as

\begin{equation}
\Pi _{A,B}(0,0)=(a,b)_{1}P(A_{2}B_{2})+(a,b)_{2}P(A_{2}\bar{B}%
_{2})+(a,b)_{3}P(\bar{A}_{2}B_{2})+(a,b)_{4}P(\bar{A}_{2}\bar{B}_{2}),
\label{00payoff}
\end{equation}%
which, as is the case for a), we express using Eqs.~(\ref{Marginals}) in
terms of the joint probability distribution $P_{A_{1},A_{2},B_{1},B_{2}}$ as

\begin{gather}
\Pi _{A,B}(0,0)=(a,b)_{1}\{P(A_{1}A_{2}B_{1}B_{2})+P(A_{1}A_{2}\bar{B}%
_{1}B_{2})+P(\bar{A}_{1}A_{2}B_{1}B_{2})+P(\bar{A}_{1}A_{2}\bar{B}%
_{1}B_{2})\}+  \notag \\
(a,b)_{2}\{P(A_{1}A_{2}B_{1}\bar{B}_{2})+P(A_{1}A_{2}\bar{B}_{1}\bar{B}%
_{2})+P(\bar{A}_{1}A_{2}B_{1}\bar{B}_{2})+P(\bar{A}_{1}A_{2}\bar{B}_{1}\bar{B%
}_{2})\}+  \notag \\
(a,b)_{3}\{P(A_{1}\bar{A}_{2}B_{1}B_{2})+P(A_{1}\bar{A}_{2}\bar{B}%
_{1}B_{2})+P(\bar{A}_{1}\bar{A}_{2}B_{1}B_{2})+P(\bar{A}_{1}\bar{A}_{2}\bar{B%
}_{1}B_{2})\}+  \notag \\
(a,b)_{4}\{P(A_{1}\bar{A}_{2}B_{1}\bar{B}_{2})+P(A_{1}\bar{A}_{2}\bar{B}_{1}%
\bar{B}_{2})+P(\bar{A}_{1}\bar{A}_{2}B_{1}\bar{B}_{2})+P(\bar{A}_{1}\bar{A}%
_{2}\bar{B}_{1}\bar{B}_{2})\}.  \label{PayoffBefore}
\end{gather}%
Observing that we assume that Bell's inequalities hold, at this stage, once
again, we refer to Fine's second theorem and insert the probability
distribution $P_{A_{1},A_{2},B_{1},B_{2}}$ given by Eqs.~(\ref{Fine1},\ref%
{Fine2}) into Eq.~(\ref{PayoffBefore}) in order to re-express it in terms of
joint probabilities $P_{A_{i},B_{j}}$ (for $i=1,2$ and $j=1,2$). This
interestingly leads to obtaining (\ref{00payoff}) again.

Note that the requirement b) states that when Bell's inequalities hold,
Alice's and Bob's payoffs at the strategy pair $(0,0)$ are $a_{4}$\ and $%
b_{4}$, respectively. To find what constraints this puts on joint
probabilities $P_{A_{i},B_{j}}$, we re-express (\ref{00payoff}) as

\begin{eqnarray}
\Pi _{A}(0,0)
&=&(a_{2}-a_{4})P(A_{2})+(a_{3}-a_{4})P(B_{2})+(a_{1}-a_{2}-a_{3}+a_{4})P(A_{2}B_{2})+a_{4},
\notag \\
\Pi _{B}(0,0)
&=&(b_{2}-b_{4})P(A_{2})+(b_{3}-b_{4})P(B_{2})+(b_{1}-b_{2}-b_{3}+b_{4})P(A_{2}B_{2})+b_{4},
\label{PayoffAfter}
\end{eqnarray}%
and then set $\Pi _{A}(0,0)=a_{4}$ and $\Pi _{B}(0,0)=b_{4}$ to obtain

\begin{equation}
P(A_{2})=0=P(B_{2}),  \label{ConstraintsA}
\end{equation}%
which defines the second set of constraints on joint probabilities.

The above analysis allows us to look forward to the situation when a joint
probability distribution $P_{A_{1},A_{2},B_{1},B_{2}}$ does not exist and/or
cannot be found from $P_{A_{i},B_{j}}$---a situation that corresponds when
Bell's inequalities are violated. We will consider this under the assumption
that the constraints on $P_{A_{i},B_{j}}$, that are obtained above and are
given by the inequalities (\ref{Constraints}), continue to hold true. This
assumption guarantees that the classical game, with its particular outcome
and the corresponding payoffs, emerges when Bell's inequalities hold and
thus the classical game remains embedded within the corresponding quantum
game.

Fine's analysis is used in above to find the constraints (\ref{ConstraintsB},%
\ref{ConstraintsA}) that we insert, in the following step, into Eqs.~(\ref%
{NEx},\ref{NEy}), that are relevant to a general strategy pair $(x^{\star
},y^{\star })$, to find if a strategy pair that is different from the
classical case of $(x^{\star },y^{\star })=(0,0)$, emerges as a NE when the
system of Bell's inequalities (\ref{BIs}) is violated and, therefore, a
probability distribution $P_{A_{1},A_{2},B_{1},B_{2}}$ cannot be found whose
marginals are $P_{A_{i},B_{j}}$. This leads us to obtain

\begin{eqnarray}
\Pi _{A}(x^{\star },y^{\star })-\Pi _{A}(x,y^{\star }) &=&(x^{\star
}-x)[y^{\star }(1-\triangle _{1}/\triangle _{2})P(B_{1})-1]\triangle
_{2}P(A_{1})\geq 0,  \notag \\
\Pi _{B}(x^{\star },y^{\star })-\Pi _{B}(x^{\star },y) &=&(y^{\star
}-y)[x^{\star }(1-\triangle _{1}/\triangle _{2})P(A_{1})-1]\triangle
_{2}P(B_{1})\geq 0.  \label{NewNE}
\end{eqnarray}%
Now, as $\triangle _{1},\triangle _{2}>0$ these inequalities once again
generate the outcome $(x^{\star },y^{\star })=(0,0)$. No new NE, therefore,
emerges for PD even when Bell's inequalities are violated and the quantum
game generates the same outcome as does the classical game.

\section{Analysis of the quantum Matching Pennies game}

The Matching Pennies (MP) game involves two players Alice and Bob and each
player has a penny that s/he secretly flips to heads $\mathcal{H}$ or tails $%
\mathcal{T}$. Players are not permitted to communicate and they disclose
their strategies to a referee who organizes the game. If the referee finds
that the two pennies match (both heads or both tails), he takes one dollar
from Bob and gives it to Alice ($+1$ for Alice, $-1$ for Bob) and if the
pennies mismatch (one heads and one tails), the referee takes one dollar
from Alice and gives it to Bob ($-1$ for Alice, $+1$ for Bob). As one
player's gain is exactly equal to the other player's loss the game is
zero-sum with the payoff matrix

\begin{equation}
\begin{array}{c}
\text{Alice}%
\end{array}%
\begin{array}{c}
\mathcal{H} \\ 
\mathcal{T}%
\end{array}%
\overset{\overset{%
\begin{array}{c}
\text{Bob}%
\end{array}%
}{%
\begin{array}{cccc}
\mathcal{H} &  &  & \mathcal{T}%
\end{array}%
}}{\left( 
\begin{array}{cc}
(+1,-1) & (-1,+1) \\ 
(-1,+1) & (+1,-1)%
\end{array}%
\right) }.  \label{Matrix}
\end{equation}

No pure strategy NE \cite{Rasmusen} exists and a unique mixed strategy NE
emerges in which both players select the strategies $\mathcal{H}$ and $%
\mathcal{T}$ with the probability of $1/2$. At the strategy pair $(x^{\star
},y^{\star })=(1/2,1/2)$ players receive $\Pi _{A}(1/2,1/2)=0=\Pi
_{B}(1/2,1/2)$.

As in the quantum game the Eq.~(\ref{payoffs}) give the players' payoff
relations, for a NE strategy pair $(x^{\star },y^{\star })$ we obtain

\begin{gather}
\Pi _{A}(x^{\star },y^{\star })-\Pi _{A}(x,y^{\star })=[y^{\star }\left\{
\Pi _{A}(S_{1},S_{1}^{\prime })-\Pi _{A}(S_{2},S_{1}^{\prime })-\Pi
_{A}(S_{1},S_{2}^{\prime })+\Pi _{A}(S_{2},S_{2}^{\prime })\right\}  \notag
\\
+\left\{ \Pi _{A}(S_{1},S_{2}^{\prime })-\Pi _{A}(S_{2},S_{2}^{\prime
})\right\} ](x^{\star }-x)\geq 0,  \notag \\
\Pi _{B}(x^{\star },y^{\star })-\Pi _{B}(x^{\star },y)=[x^{\star }\left\{
\Pi _{B}(S_{1},S_{1}^{\prime })-\Pi _{B}(S_{1},S_{2}^{\prime })-\Pi
_{B}(S_{2},S_{1}^{\prime })+\Pi _{B}(S_{2},S_{2}^{\prime })\right\}  \notag
\\
+\left\{ \Pi _{B}(S_{2},S_{1}^{\prime })-\Pi _{B}(S_{2},S_{2}^{\prime
})\right\} ](y^{\star }-y)\geq 0,  \label{QNE}
\end{gather}%
where Eqs.~(\ref{payoffs-parts}) and the matrix (\ref{Matrix}) gives

\begin{eqnarray}
\Pi _{A}(S_{1},S_{1}^{\prime }) &=&P(A_{1}B_{1})-P(A_{1}\bar{B}_{1})-P(\bar{A%
}_{1}B_{1})+P(\bar{A}_{1}\bar{B}_{1})=-\Pi _{B}(S_{1},S_{1}^{\prime }), 
\notag \\
\Pi _{A}(S_{1},S_{2}^{\prime }) &=&P(A_{1}B_{2})-P(A_{1}\bar{B}_{2})-P(\bar{A%
}_{1}B_{2})+P(\bar{A}_{1}\bar{B}_{2})=-\Pi _{B}(S_{1},S_{2}^{\prime }), 
\notag \\
\Pi _{A}(S_{2},S_{1}^{\prime }) &=&P(A_{2}B_{1})-P(A_{2}\bar{B}_{1})-P(\bar{A%
}_{2}B_{1})+P(\bar{A}_{2}\bar{B}_{1})=-\Pi _{B}(S_{2},S_{1}^{\prime }), 
\notag \\
\Pi _{A}(S_{2},S_{2}^{\prime }) &=&P(A_{2}B_{2})-P(A_{2}\bar{B}_{2})-P(\bar{A%
}_{2}B_{2})+P(\bar{A}_{2}\bar{B}_{2})=-\Pi _{B}(S_{2},S_{2}^{\prime }).
\label{QPayoffsPartsExplicit}
\end{eqnarray}%
The right sides of these Equations express the fact that, as it is the case
with the classical game, the quantum game is also zero-sum game.

As it was the case for the PD game, we define

\begin{equation}
\triangle _{x^{\star }=1/2}=\Pi _{A}(1/2,1/2)-\Pi _{A}(x,1/2)\text{, \ \ }%
\triangle _{y^{\star }=1/2}=\Pi _{B}(1/2,1/2)-\Pi _{B}(1/2,y),
\end{equation}%
and insert from Eqs.~(\ref{QPayoffsPartsExplicit}) into Eqs.~(\ref{QNE}) to
obtain

\begin{gather}
\triangle _{x^{\star }=1/2}=(1/2)[P(A_{1}B_{1})-P(A_{1}\bar{B}_{1})-P(\bar{A}%
_{1}B_{1})+P(\bar{A}_{1}\bar{B}_{1})  \notag \\
-P(A_{2}B_{1})+P(A_{2}\bar{B}_{1})+P(\bar{A}_{2}B_{1})-P(\bar{A}_{2}\bar{B}%
_{1})  \notag \\
P(A_{1}B_{2})-P(A_{1}\bar{B}_{2})-P(\bar{A}_{1}B_{2})+P(\bar{A}_{1}\bar{B}%
_{2})  \notag \\
-P(A_{2}B_{2})+P(A_{2}\bar{B}_{2})+P(\bar{A}_{2}B_{2})-P(\bar{A}_{2}\bar{B}%
_{2})](1/2-x)\geq 0,  \label{MPNEx}
\end{gather}%
and

\begin{gather}
\triangle _{y^{\star }=1/2}=(1/2)[-P(A_{1}B_{1})+P(A_{1}\bar{B}_{1})+P(\bar{A%
}_{1}B_{1})-P(\bar{A}_{1}\bar{B}_{1})  \notag \\
P(A_{1}B_{2})-P(A_{1}\bar{B}_{2})-P(\bar{A}_{1}B_{2})+P(\bar{A}_{1}\bar{B}%
_{2})  \notag \\
-P(A_{2}B_{1})+P(A_{2}\bar{B}_{1})+P(\bar{A}_{2}B_{1})-P(\bar{A}_{2}\bar{B}%
_{1})  \notag \\
P(A_{2}B_{2})-P(A_{2}\bar{B}_{2})-P(\bar{A}_{2}B_{2})+P(\bar{A}_{2}\bar{B}%
_{2})](1/2-y)\geq 0.  \label{MPNEy}
\end{gather}

We now insert from Eqs.~(\ref{Marginals}) into Eqs.~(\ref{MPNEx}, \ref{MPNEy}%
) to obtain

\begin{eqnarray}
\triangle _{x^{\star }=1/2} &=&2[P(A_{1}\bar{A}_{2}B_{1}B_{2})-P(A_{1}\bar{A}%
_{2}\bar{B}_{1}\bar{B}_{2})-P(\bar{A}_{1}A_{2}B_{1}B_{2})+P(\bar{A}_{1}A_{2}%
\bar{B}_{1}\bar{B}_{2})](1/2-x),  \notag \\
&&  \label{MPNE1} \\
\triangle _{y^{\star }=1/2} &=&2[P(A_{1}A_{2}\bar{B}%
_{1}B_{2})-P(A_{1}A_{2}B_{1}\bar{B}_{2})-P(\bar{A}_{1}\bar{A}_{2}\bar{B}%
_{1}B_{2})+P(\bar{A}_{1}\bar{A}_{2}B_{1}\bar{B}_{2})](1/2-y),  \notag \\
&&  \label{MPNE2}
\end{eqnarray}%
that express $\triangle _{x^{\star }=1/2}$ and $\triangle _{y^{\star }=1/2}$
in terms of the joint probability distribution $P_{A_{1},A_{2},B_{1},B_{2}}$%
. As it was the case with the PD game, at this stage we demand that the
strategy pair $(1/2,1/2)$ results as a NE when players share a classical
physical system, for which the system (\ref{BIs}) of Bell's inequalities
hold and the joint probability distribution $P_{A_{1},A_{2},B_{1},B_{2}}$
can be found from $P_{A_{i},B_{j}}$ using Fine's second theorem. It,
therefore, seems natural to impose the following two requirements: When the
system of Bell's inequalities (\ref{BIs})\ hold we have: a) Nash
inequalities for the strategy pair $(1/2,1/2)$ hold i.e. $\triangle
_{x^{\star }=1/2}\geq 0$, $\triangle _{y^{\star }=1/2}\geq 0$, b) the
payoffs for players Alice and Bob at the strategy pair $(1/2,1/2)$ are zero
both.

As has been the case with the PD game, a keen reader may ask here why the
converse situation is not adopted: to require that if $\triangle _{x^{\star
}=1/2}\geq 0$ and $\triangle _{y^{\star }=1/2}\geq 0$ then the system of
Bell's inequalities holds. As the violation of the system of Bell's
inequalities does not necessarily lead to a new non-classical NE, and we
well may have $\triangle _{x^{\star }=1/2}\geq 0$ and $\triangle _{y^{\star
}=1/2}\geq 0$ even when Bell's inequalities are violated, we consider it not
to be a valid requirement.

To address a) we require that if the probability distribution $%
P_{A_{1},A_{2},B_{1},B_{2}}$, obtained from Fine's second theorem, is
inserted in Eqs.~(\ref{MPNE1},\ref{MPNE2}) then we have $\triangle
_{x^{\star }=1/2}\geq 0$ and $\triangle _{y^{\star }=1/2}\geq 0$. Inserting
from Eqs.~(\ref{Fine1},\ref{Fine2}) to Eqs.~(\ref{MPNE1},\ref{MPNE2})
reduces them to simpler form:

\begin{eqnarray}
\triangle _{x^{\star }=1/2}
&=&2[P(A_{2})-P(A_{1})][1-P(B_{1})-P(B_{2})](1/2-x)\geq 0,  \notag \\
\triangle _{y^{\star }=1/2}
&=&2[P(B_{1})-P(B_{2})][1-P(A_{1})-P(A_{2})](1/2-y)\geq 0,
\end{eqnarray}%
which defines the first set of constraints on joint probabilities as

\begin{eqnarray}
\lbrack P(A_{2})-P(A_{1})][1-P(B_{1})-P(B_{2})] &=&0,  \label{MPConstraints1}
\\
\lbrack P(B_{1})-P(B_{2})][1-P(A_{1})-P(A_{2})] &=&0.  \label{MPConstraints2}
\end{eqnarray}

Now we translate the requirement b) into the second set of constraints on
joint probabilities. Consider the relations (\ref{payoffs})\ to find
players' payoffs at the strategy pair $(1/2,1/2)$ as

\begin{equation}
\Pi _{A,B}(1/2,1/2)=(1/4)\{\Pi _{A,B}(S_{1},S_{1}^{\prime })+\Pi
_{A,B}(S_{1},S_{2}^{\prime })+\Pi _{A,B}(S_{2},S_{1}^{\prime })+\Pi
_{A,B}(S_{2},S_{2}^{\prime })\},  \label{MPNEpayoffs1}
\end{equation}%
where

\begin{eqnarray}
\Pi _{A}(S_{1},S_{1}^{\prime }) &=&P(A_{1}B_{1})-P(A_{1}\bar{B}_{1})-P(\bar{A%
}_{1}B_{1})+P(\bar{A}_{1}\bar{B}_{1})=-\Pi _{B}(S_{1},S_{1}^{\prime }), 
\notag \\
\Pi _{A}(S_{1},S_{2}^{\prime }) &=&P(A_{1}B_{2})-P(A_{1}\bar{B}_{2})-P(\bar{A%
}_{1}B_{2})+P(\bar{A}_{1}\bar{B}_{2})=-\Pi _{B}(S_{1},S_{2}^{\prime }), 
\notag \\
\Pi _{A}(S_{2},S_{1}^{\prime }) &=&P(A_{2}B_{1})-P(A_{2}\bar{B}_{1})-P(\bar{A%
}_{2}B_{1})+P(\bar{A}_{2}\bar{B}_{1})=-\Pi _{B}(S_{2},S_{1}^{\prime }), 
\notag \\
\Pi _{A}(S_{2},S_{2}^{\prime }) &=&P(A_{2}B_{2})-P(A_{2}\bar{B}_{2})-P(\bar{A%
}_{2}B_{2})+P(\bar{A}_{2}\bar{B}_{2})=-\Pi _{B}(S_{2},S_{2}^{\prime }). 
\notag \\
&&  \label{MPNEpayoffs2}
\end{eqnarray}%
Inserting (\ref{MPNEpayoffs2}) in (\ref{MPNEpayoffs1}) gives

\begin{gather}
\Pi
_{A}(1/2,1/2)=(1/4)[%
\{P(A_{1}B_{1})+P(A_{1}B_{2})+P(A_{2}B_{1})+P(A_{2}B_{2})+P(\bar{A}_{1}\bar{B%
}_{1})+  \notag \\
P(\bar{A}_{1}\bar{B}_{2})+P(\bar{A}_{2}\bar{B}_{1})+P(\bar{A}_{2}\bar{B}%
_{2})\}-\{P(A_{1}\bar{B}_{1})+P(A_{1}\bar{B}_{2})+P(A_{2}\bar{B}_{1})+ 
\notag \\
P(A_{2}\bar{B}_{2})+P(\bar{A}_{1}B_{1})+P(\bar{A}_{1}B_{2})+P(\bar{A}%
_{2}B_{1})+P(\bar{A}_{2}B_{2})\}]=-\Pi _{B}(1/2,1/2).  \label{H-Hpayoff}
\end{gather}%
We now insert from Eqs.~(\ref{Marginals}) into the payoff (\ref{H-Hpayoff})
in order to express it in terms of the joint probability distribution $%
P_{A_{1},A_{2},B_{1},B_{2}}$ to obtain

\begin{equation}
\Pi _{A}(1/2,1/2)=P(A_{1}A_{2}B_{1}B_{2})-P(A_{1}A_{2}\bar{B}_{1}\bar{B}%
_{2})-P(\bar{A}_{1}\bar{A}_{2}B_{1}B_{2})+P(\bar{A}_{1}\bar{A}_{2}\bar{B}_{1}%
\bar{B}_{2}).  \label{H-Hpayoff2}
\end{equation}

As Bell's inequalities are assumed to hold, at this stage we refer to Fine's
second theorem and insert the probability distribution $%
P_{A_{1},A_{2},B_{1},B_{2}}$ given by Eqs.~(\ref{Fine1},\ref{Fine2}) into
Eq.~(\ref{H-Hpayoff2}) in order to re-express it in terms of joint
probabilities:

\begin{equation}
\Pi _{A}(1/2,1/2)=[P(B_{1})+P(B_{2})][P(A_{1})+P(A_{2})-1]-[(1+\beta
)P(A_{1})+(1-\alpha )P(A_{2})].  \label{H-Hpayoff3}
\end{equation}%
Now the requirement b) states that when Bell's inequalities hold, Alice's
and Bob's payoffs for the strategy pair $(1/2,1/2)$ are both zero, giving us
the second constraint on joint probabilities:

\begin{equation}
\lbrack P(B_{1})+P(B_{2})][P(A_{1})+P(A_{2})-1]=[(1+\beta
)P(A_{1})+(1-\alpha )P(A_{2})].  \label{MPConstraints3}
\end{equation}

After some manipulation, the constraints (\ref{MPConstraints1},\ref%
{MPConstraints2},\ref{MPConstraints3}) can be re-expressed as

\begin{eqnarray}
P(A_{1}B_{2}) &=&[(2+\beta )P(A_{1})+P(B_{1})+P(B_{2})-\alpha
P(A_{2})]/2-P(A_{1}B_{1}),  \label{ConstA} \\
P(A_{2}B_{1}) &=&[(1-\alpha )P(A_{2})+2P(B_{1})+(1+\beta
)P(A_{1})-2P(A_{1}B_{1})]/2,  \label{ConstB} \\
P(A_{2}B_{2}) &=&[2P(A_{1}B_{1})+P(A_{2})+P(B_{2})-P(A_{1})-P(B_{1})]/2.
\label{ConstC}
\end{eqnarray}

To find if the violation of Bell's inequalities may lead to a NE, which is
different from the classical NE of $(x^{\star },y^{\star })=(1/2,1/2)$, we
consider Eqs.~(\ref{QPayoffsPartsExplicit}) to obtain

\begin{gather}
\Pi _{A}(S_{1},S_{1}^{\prime })-\Pi _{A}(S_{2},S_{1}^{\prime })-\Pi
_{A}(S_{1},S_{2}^{\prime })+  \notag \\
\Pi _{A}(S_{2},S_{2}^{\prime })=4[P(A_{1}B_{1})-P(A_{2}B_{1})-P(A_{1}B_{2})],
\notag \\
\Pi _{A}(S_{1},S_{2}^{\prime })-\Pi _{A}(S_{2},S_{2}^{\prime
})=2\{2[P(A_{1}B_{2})-P(A_{2}B_{2})]+  \notag \\
\lbrack P(A_{2})-P(A_{1})]\}.  \label{DiffA}
\end{gather}

We now substitute the constraints given by Eqs.~(\ref{ConstA},\ref{ConstB},%
\ref{ConstC}) into (\ref{DiffA}) to obtain

\begin{gather}
\Pi _{A}(S_{1},S_{1}^{\prime })-\Pi _{A}(S_{2},S_{1}^{\prime })-\Pi
_{A}(S_{1},S_{2}^{\prime })+  \notag \\
\Pi _{A}(S_{2},S_{2}^{\prime })=4[4P(A_{1}B_{1})-(2+\beta )P(A_{1})+\alpha
P(A_{2})-2P(B_{1})],  \notag \\
\Pi _{A}(S_{1},S_{2}^{\prime })-\Pi _{A}(S_{2},S_{2}^{\prime })=2[(2+\beta
)P(A_{1})-4P(A_{1}B_{1})+2P(B_{1})-\alpha P(A_{2})].
\end{gather}%
This allows to write the two inequalities in (\ref{QNE}) as

\begin{eqnarray}
\Pi _{A}(x^{\star },y^{\star })-\Pi _{A}(x,y^{\star }) &=&4\Omega (y^{\star
}-1/2)(x^{\star }-x)\geq 0,  \label{QNE1} \\
\Pi _{B}(x^{\star },y^{\star })-\Pi _{B}(x^{\star },y) &=&-4\Omega (x^{\star
}-1/2)(y^{\star }-y)\geq 0,  \label{QNE2}
\end{eqnarray}%
where

\begin{equation}
\Omega =4P(A_{1}B_{1})-(2+\beta )P(A_{1})+\alpha P(A_{2})-2P(B_{1}).
\end{equation}%
Now, the inequalities (\ref{QNE1},\ref{QNE2}) state that although the
strategy pair $(x^{\star },y^{\star })=(1/2,1/2)$ remains a NE also in the
quantum game, for which the system (\ref{BIs}) of Bell's inequalities is
violated, any pair of strategies will become a NE when $\Omega =0$.

Cereceda reports in Ref.~\cite{Cereceda} that, corresponding to a maximally
entangled bipartite state, there exist two sets of joint probabilities,
which maximally violate the CHSH sum of correlations while satisfying the
constraints (\ref{normalization},\ref{locality}) that are given by
normalization and causal communication constraint. To define these
probability sets Cereceda divides the $16$ joint probabilities into two sets:

\begin{eqnarray}
\nu &=&\left\{ P(A_{1}\bar{B}_{1}),P(\bar{A}_{1}B_{1}),P(A_{1}\bar{B}_{2}),P(%
\bar{A}_{1}B_{2}),P(A_{2}\bar{B}_{1}),P(\bar{A}_{2}B_{1}),P(A_{2}B_{2}),P(%
\bar{A}_{2}\bar{B}_{2})\right\} ,  \notag \\
\mu &=&\left\{ P(A_{1}B_{1}),P(\bar{A}_{1}\bar{B}_{1}),P(A_{1}B_{2}),P(\bar{A%
}_{1}\bar{B}_{2}),P(A_{2}B_{1}),P(\bar{A}_{2}\bar{B}_{1}),P(A_{2}\bar{B}%
_{2}),P(\bar{A}_{2}B_{2})\right\} .  \notag \\
&&
\end{eqnarray}%
In terms of these the first probability set is then given as

\begin{equation}
P_{l}=(2+\sqrt{2})/8\text{ for all }P_{l}\in \mu ,\text{ and }P_{m}=(2-\sqrt{%
2})/8\text{ for all }P_{m}\in \nu ,  \label{First probability set}
\end{equation}%
whereas the second probability set is given as

\begin{equation}
P_{l}=(2-\sqrt{2})/8\text{ for all }P_{l}\in \mu ,\text{ and }P_{m}=(2+\sqrt{%
2})/8\text{ for all }P_{m}\in \nu \text{.}  \label{Second probability set}
\end{equation}%
These two sets, while being consistent with the normalization and causal
communication constraints given by Eqs. (\ref{normalization},\ref{locality}%
), provide the maximum absolute limit of $2\sqrt{2}$ for the CHSH sum of
correlations.

To evaluate $\Omega $ for these two probability sets we use the definition (%
\ref{Gamma}) to find $\gamma $ for $n=1,2$ and $m\neq k=1,2$. We, therefore,
consider the quantities

\begin{eqnarray}
&&[P(A_{1}B_{1})+P(B_{2})-P(A_{1}B_{2}),P(B_{1}),P(B_{2})],  \notag \\
&&[P(A_{1}B_{2})+P(B_{1})-P(A_{1}B_{1}),P(B_{2}),P(B_{1})],  \notag \\
&&[P(A_{2}B_{1})+P(B_{2})-P(A_{2}B_{2}),P(B_{1}),P(B_{2})],  \notag \\
&&[P(A_{2}B_{2})+P(B_{1})-P(A_{2}B_{1}),P(B_{2}),P(B_{1})].
\label{expressions}
\end{eqnarray}%
The first of which, for example, is expressed as

\begin{eqnarray}
&&\{[P(A_{1}B_{1})+P(A_{1}\bar{B}_{1})+P(A_{1}B_{2})+P(A_{1}\bar{B}%
_{2})][P(A_{1}B_{1})+P(\bar{A}_{1}B_{1})+  \notag \\
&&P(A_{2}B_{1})+P(\bar{A}_{2}B_{1})]+[P(A_{1}B_{2})+P(\bar{A}%
_{1}B_{2})+P(A_{2}B_{2})+P(\bar{A}_{2}B_{2})]-  \notag \\
&&[P(A_{1}B_{1})+P(A_{1}\bar{B}_{1})+P(A_{1}B_{2})+P(A_{1}\bar{B}%
_{2})][P(A_{1}B_{2})+P(\bar{A}_{1}B_{2})+  \notag \\
&&P(A_{2}B_{2})+P(\bar{A}_{2}B_{2})],[P(A_{1}B_{1})+P(\bar{A}%
_{1}B_{1})+P(A_{2}B_{1})+P(\bar{A}_{2}B_{1})],  \notag \\
&&[P(A_{1}B_{2})+P(\bar{A}_{1}B_{2})+P(A_{2}B_{2})+P(\bar{A}_{2}B_{2})]\},
\end{eqnarray}%
which reduces itself to $\{1,1,1\}$ for the probability set (\ref{First
probability set}). The same is the case with the remaining expressions of (%
\ref{expressions}) for this probability set. So we obtain $\gamma =1$ that
gives $\Omega =0$. Similarly, for the probability set (\ref{Second
probability set}) we also obtain $\Omega =0$. In view of the Inequalities (%
\ref{QNE1},\ref{QNE2}), in the quantum MP game with players sharing a
maximally entangled state to play the game, this results in any strategy set 
$(x^{\star },y^{\star })$ being a NE. As for $\gamma =1$ we obtain $\alpha
=P(A_{1})$ and $\beta =P(A_{2})$ from Eq.~(\ref{AlphaBeta}), the constraints
(\ref{ConstA},\ref{ConstB},\ref{ConstC}) are satisfied for both the
probability sets in Eqs. (\ref{First probability set},\ref{Second
probability set}).

\section{Discussion}

As the present paper builds up on our earlier work that constructs quantum
games from non-factorizable joint probabilities \cite{IqbalCheon}, its brief
review is in order. That work also uses the setting of a quantum correlation
experiment and players's strategies are classical as is the case in the
present approach. A classical game is re-expressed in terms of factorizable
joint probabilities relevant to a shared physical system under the
assumption that factorizable joint probabilities correspond to classicality.
It is found that by requiring a classical outcome of the game to emerge for
factorizable joint probabilities results in constraints on the joint
probabilities. These constraints ensure that the classical game
corresponding to factorizable joint probabilities remains a subset of the
quantum game. Retaining these constraints and allowing the joint
probabilities to become non-factorizable leads to the corresponding quantum
game. When played in this setting, it is found that new quantum mechanical
NE emerge, for instance, for the game of Matching Pennies \cite{IqbalAbbott}
that correspond, interestingly, to the sets of joint probabilities that
maximally violate the CHSH inequality \cite{Peres}.

The relation, however, which this approach establishes between the
classicality of the shared physical system, as expressed by a system of
Bell's inequalities, and a classical game is not straightforward. It is
because Bell's inequalities may not be violated even when the corresponding
set of joint probabilities is non-factorizable i.e. non-factorizability is
necessary but not sufficient to violate Bell's inequalities. A suggested
explanation can be to state that such a game resides in the so-called
pseudo-classical domain, where Bell's inequalities are not violated, and
where a quantum game is treated as if players are simultaneously playing
several classical games \cite{CheonTsutsui}.

The present paper introduces a new approach to quantize a two-player
two-strategy game. This approach, once again, uses the setting of quantum
correlation experiments in which players' strategies remain classical. The
quantum game is now obtained from the non-classical feature of the shared
physical system consisting of the violation of Bell's inequalities. This
situation corresponds when a joint probability distribution $%
P_{A_{1},A_{2},B_{1},B_{2}}$ does not exist, whose marginals are the joint
probabilities $P_{A_{i},B_{j}}$.

The argument presented in this paper can be described as follows. We begin
by putting a classical game into a suitable format that permits us to
consider the situation when a joint probability distribution $%
P_{A_{1},A_{2},B_{1},B_{2}}$, whose marginals are the joint probabilities $%
P_{A_{i},B_{j}}$, does not exist. We obtain this format by expressing
players' payoff relations in terms of the joint probabilities $%
P_{A_{i},B_{j}}$. With the payoff relations expressed in this way we select
an arbitrary strategy pair $(x^{\star },y^{\star })$ and find constraints on
the joint probabilities $P_{A_{i},B_{j}}$ that produce this NE and its
corresponding classical payoffs to the players. Assuming that a joint
probability distribution $P_{A_{1},A_{2},B_{1},B_{2}}$ exists that can be
found from Fine's second theorem, we re-express the obtained constraints in
terms of the joint probability distribution. We take the strategy pair $%
(x^{\star },y^{\star })$ to be the NE of the classical game, and players'
payoffs for this strategy pair to be their payoffs in the classical game.
This allows us to use Eqs.~(\ref{Fine1},\ref{Fine2}) and to express obtained
constraints on the joint probability distribution $%
P_{A_{1},A_{2},B_{1},B_{2}}$ as constraints on joint probabilities $%
P_{A_{i},B_{j}}$. The obtained constraints ensure that when the system of
Bell's inequalities holds we obtain the classical NE as the outcome of the
game. The quantum game is then obtained by retaining these constraints while
allowing the joint probability distribution $P_{A_{1},A_{2},B_{1},B_{2}}$
not to exist. Considering a particular game we then investigate if this
leads to NE that are non-classical.

We investigate games of Prisoners' Dilemma and Matching Pennies both of
which have been studied earlier using other quantization schemes \cite%
{EWL,IqbalCheon,IqbalAbbott}. For PD we find that when the system of Bell's
inequalities does not hold, and a joint probability distribution $%
P_{A_{1},A_{2},B_{1},B_{2}}$ does not exist, this cannot change or shift the
classical NE of the game. For this game an identical result was reported in
Ref.~\cite{IqbalCheon} using non-factorizable joint probabilities. For MP we
find that in this quantization scheme the classical NE remains intact even
when the system (\ref{BIs}) of Bell's inequalities is violated and the joint
probability distribution $P_{A_{1},A_{2},B_{1},B_{2}}$ does not exist.
However, when a maximally entangled state is shared between players, any
pair of strategies becomes a NE. This result diverges away from the one
obtained earlier \cite{IqbalAbbott} using non-factorizable joint
probabilities. We believe it is because non-factorizability is not
equivalent to a joint probability distribution $P_{A_{1},A_{2},B_{1},B_{2}}$
not existing---a situation that motivates this paper. This analysis confirms
that an outcome of a quantum game is dependent on the quantization route
taken.

The results obtained show that for quantum games played in the setting of
quantum correlation experiments the sharing of quantum resources does not
always lead to players doing better than what they can do in the classical
game. Players sharing a quantum system (for which Bell's inequalities are
violated) do not automatically become better off relative to the ones who
share classical system in order to physically implement the same game. This
is observed to be the case with the game of PD. On the contrary, for the
game of MP the situation becomes quite different as players' sharing of the
quantum system (that corresponds to a maximally entangled state) leads to
the situation of any pair of strategies existing as NE. It, therefore, shows
that the consequence of sharing of a quantum system depends on the
particular original classical game the players play. This is in agreement
with the results reported earlier by Shimamura et al. \cite{Shimamura1}
showing that within Eisert et al.'s quantization scheme \cite{EWL} certain
quantum states cannot be used to quantize certain classical games as by
doing so classical results cannot be reproduced. Our results convey the same
message though the quantum games we consider are constructed directly from
Bell's inequalities.

An interesting question is to ask about the consequence the new quantum
solutions have regarding the original considered game. We believe that the
new quantum solutions have a consequence regarding the original considered
game only when a classical game emerges due to classicality of the shared
physical system---in the sense that when the shared physical system does not
violate Bell's inequalities the resulting game attains a classical
interpretation (both in terms of the players' payoffs and the resulting pair
of strategies defining the NE).

Some of the known criticisms of quantum games can be stated as follows: a) a
quantum game is an ad-hoc construction that does not teach us anything new
about quantum mechanics, b) it is known in game theory that if you change
the rules of an old game, you get a new game. The fact that new NE appear in
the quantum game is not surprising and does not imply anything about the old
game, c) by including the new `quantum' moves in a pay-off matrix, one can
reformulate the quantum game as a purely classical game \cite{EnkPike} with
players having access to an extended set of available pure strategies.

In reply to a) we state that a quantum game offers a reasonable way to
extend a classical game towards quantum domain and such an extension cannot,
and should not, be assumed to open new avenues for quantum mechanics.
However, it is the game theory for which a new avenue is opened in that
taking a game to the quantum regime is found to have consequences for the
outcome the considered game. In reply to b) we state that any construction
of a quantum game is an extension of the original game and therefore the
rules for playing this extended game also need to take into account the
particular extension made. The rules of the extended game are relevant to
the new game and therefore cannot be expected to remain identical to the
ones in the original game. Under reasonable conditions, however, the
extended game and its rules, should be reducible to the original game. In
the construction we develop here, the reasonable constraints are a system of
Bell's inequalities. In reply to c) we state that as in the present
construction a quantum game corresponds only when Bell's inequalities,
relevant to the shared physical system, are violated, and that the classical
game corresponds when this system holds. As players' strategies remain
classical in the quantum game, this offers the closest situation without
changing the rules of the game---especially when compared to the usual case
in which players' allowed strategy sets are extended to unitary
transformations.

In the setting we use, the players strategies become dependent on the shared
physical system via the payoff relations. For a shared system that exhibits
quantum correlations, this may give the impression that the game is changed
from a non-cooperative game to a cooperative game with different rules. We
agree that it is possible to model our quantum game by introducing, for
instance, pre-play negotiations into the standard setting of a
non-cooperative game. Our objective, however, is different in that we ask
how the violation of Bell's inequalities by the shared physical system
impacts the outcome of a non-cooperative game. The possibility of modeling
this impact by introducing some kind of cooperation between the players
cannot be equated to changing a non-cooperative game into an explicitly
cooperative game.

Agreeing with the earlier reported results \cite%
{BenjaminHaydenC,BenjaminHayden,Johnson,Shimamura,Shimamura2,Mendes} this
paper shows that players sharing a quantum mechanical system may or may
result in new outcomes as this depends on the original classical game.
Secondly, we present the first analysis that directly exploits a system of
Bell's inequalities, together with Fine's results, establishing a direct
link between a system of Bell's inequalities and the existence of a joint
probability distribution in the construction of quantum games. Possible
directions for further investigation may include expressing the outcome(s)
of a quantum game in terms of the amount of the violation of Bell's
inequalities such that they are reduced to classical outcomes when there
exists no violation. We believe the quantization approach proposed in this
paper can be extended to multi-player games if Fine's results could be
accordingly extended to more than $4$ bivalent observables.

\textbf{Acknowledgment:} This project was supported at the University of
Adelaide by the Australian Research Council under the Discovery Projects
scheme (Grant No. DP0771453).

\end{document}